\documentclass[a4paper,10pt,twoside]{just}

\usepackage{multicol}
\usepackage{graphicx}
\usepackage{booktabs}
\usepackage{amssymb,bm,mathrsfs,bbm,amscd}
\usepackage[tbtags]{amsmath}
\usepackage{lastpage}
\usepackage{CJK}

\begin{document}
\begin{CJK*}{GBK}{song}

\fancyhead[c]{\small  10th International Workshop on $e^+e^-$ collisions from $\phi$ to $\psi$ (PhiPsi15)}
 \fancyfoot[C]{\small PhiPsi15-\thepage}

\footnotetext[0]{Received 14 Sep. 2014}

\title{Towards nature of  the X(3872) resonance\thanks{Supported in part by RFBR, Grant No 13-02-00039,
and Interdisciplinary project No 102 of Siberian division of RAS.
}}

\author{%
      N.N. Achasov$^{1;1)}$\email{achasov@math.nsc.ru}
\quad E.V. Rogozina$^{2}$} \maketitle

\address{%
$^1$ Laboratory of Theoretical Physics, Sobolev Institute for
Mathematics,  630090, Novosibirsk, Russian Federation\\
 $^2$ Novosibirsk State University, 630090, Novosibirsk, Russian Federation\\ }

\begin{abstract}
We construct  spectra of decays of the resonance $X(3872)$  with
good analytical and unitary properties which allows to define the
branching ratio of the $X(3872) \to D^{*0}\bar D^0 + c.c.$ decay
studying only one more decay, for example, the
$X(3872)\to\pi^+\pi^- J/\psi(1S)$ decay, and  show that our
spectra are  effective means of selection of models for the
resonance $X(3872)$.

 Then we discuss the scenario where the
$X(3872)$ resonance is the
 $c\bar c = \chi_{c1}(2P)$ charmonium which "sits on" the  $D^{*0}\bar D^0$
 threshold.

 We explain the shift of  the mass of the
 $X(3872)$ resonance with respect to the prediction of a potential model for the mass of the
$\chi_{c1}(2P)$ charmonium by the contribution of the virtual
$D^*\bar D+c.c.$ intermediate states into the self energy of the
$X(3872)$ resonance. This allows us to estimate the coupling
constant of the $X(7872)$ resonance with the $D^{*0}\bar D^0$
channel, the branching ratio of the $X(3872) \to D^{*0}\bar D^0 +
c.c.$ decay, and the branching ratio of  the $X(3872)$ decay into
all non-$D^{*0}\bar D^0 + c.c.$ states. We predict a significant
number of unknown decays of    $X(3872)$ via two gluons:
$X(3872)\to gluon\ gluon\to hadrons$.

\end{abstract}



\begin{multicols}{2}

\section{Introduction}

The $X(3872)$ resonance became the first in discovery of the
resonant structures $XYZ$ ($X(3872)$, $Y(4260)$, $Z_b^+(10610)$,
$Z_b^+(10650)$, $Z_c^+(3900)$), the interpretations of which as
hadron states assumes existence in them at least pair of heavy and
pair of light quarks in this or that form.

 Thousands of articles
on this subject already were published in spite of the fact that
many properties of new resonant structures are not defined yet and
not all possible mechanisms of dynamic generation of these
structures are studied, in particular, the role of  the anomalous
Landau thresholds is not studied.

 Anyway, this spectroscopy took the
central place in physics of hadrons.

 Below we give reasons that
$X(3872)$, $I^G(J^{PC})=0^+(1^{++})$, is the $\chi_{c1}(2P)$
charmonium and suggest a physically clear program of experimental
researches for verification of our assumption.

\section{ \bf\boldmath{$ $} How to learn the branching ratio $X(3872)
\to D^{*0}\bar D^0 + c.c.$ \cite{NNAEVR}}
 The mass spectrum  $\pi^+\pi^-J/\psi(1S)$ looks as the ideal
Breit-Wigner one   in the $X(3872)\to\pi^+\pi^-J/\psi(1S)$ decay,
see Fig. \ref{fig1} (a).
\end{multicols}
\ruleup
\begin{center}
\begin{tabular}{ccc}
\includegraphics[width=8.5cm,height=5.9cm]{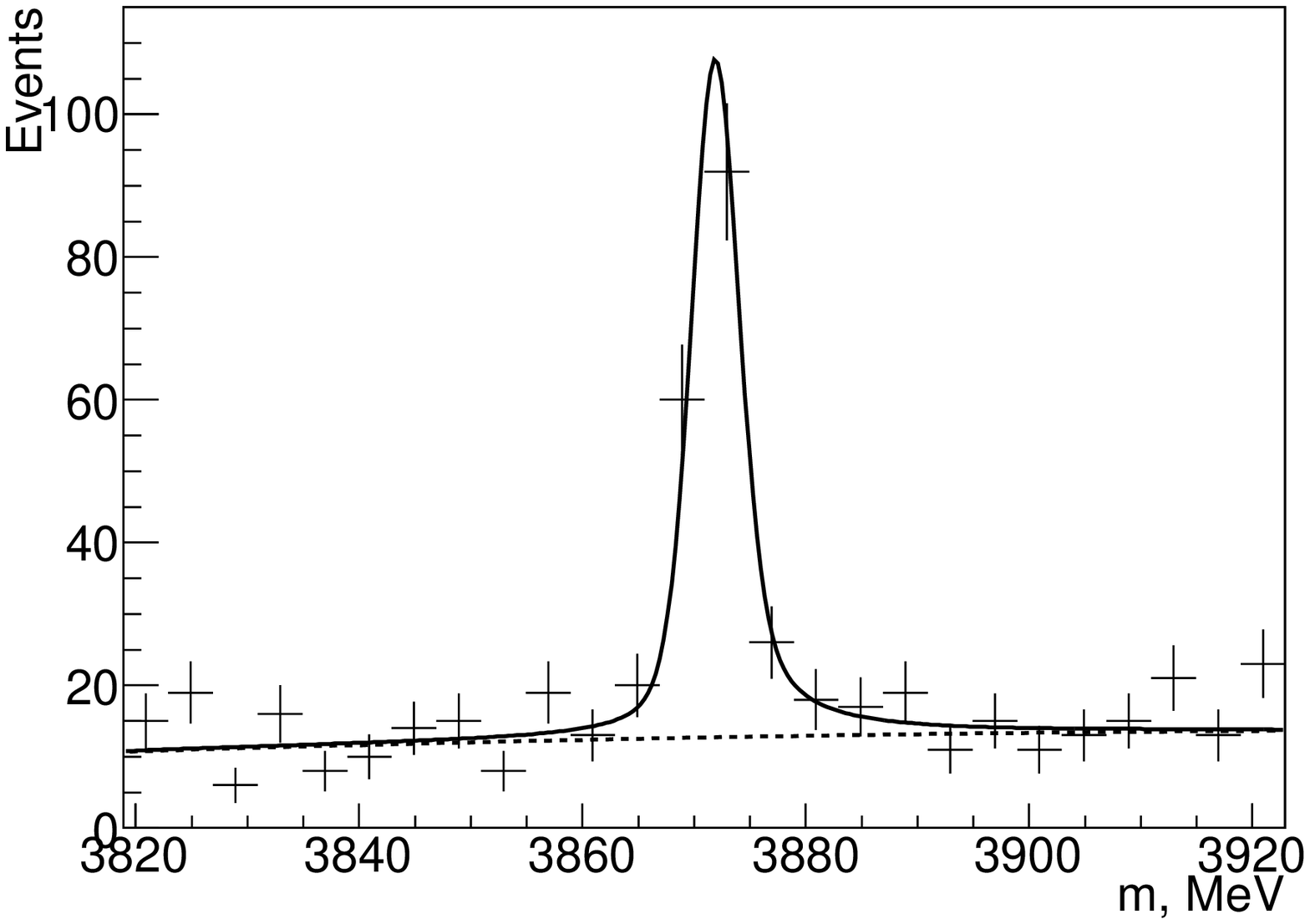}& \includegraphics[width=8.5cm,height=5.9cm]{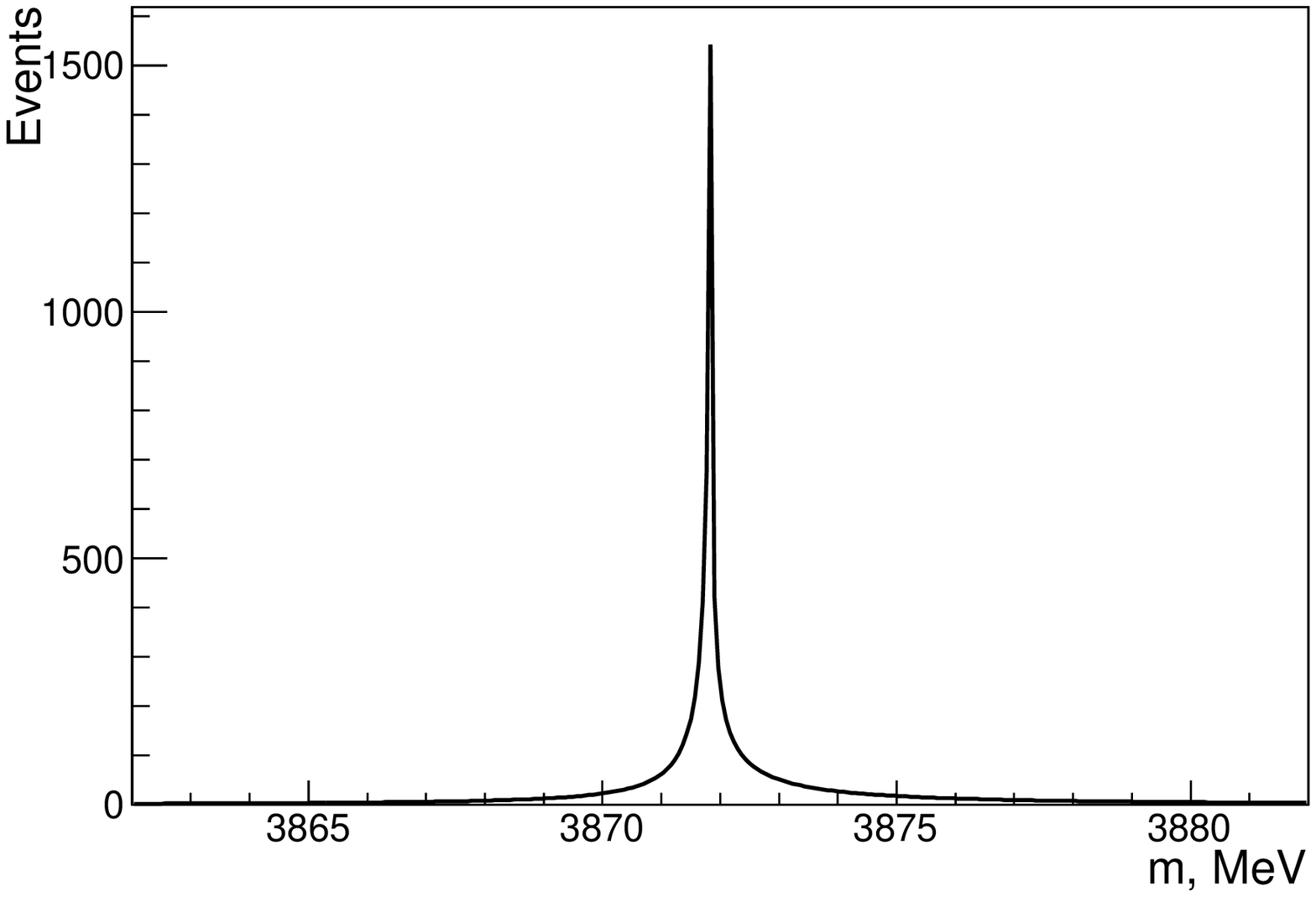}\\ (a)&(b)
\end{tabular}
\figcaption{(a) The Belle data \cite{Belle11} on the invariant
$\pi^+\pi^- J/\psi(1S)$ mass ($m$) distribution. The solid line is
our theoretical one with taking into account the Belle energy
resolution. The dotted line is second-order polynomial for the
incoherent background. (b) Our undressed theoretical
line.\label{fig1} }
\end{center}
\ruledown

\begin{multicols}{2}

 The mass spectrum $\pi^+\pi^-\pi^0 J/\psi(1S)$ in
the $X(3872)\to\pi^+\pi^-\pi^0 J/\psi(1S)$ decay
 looks in a similar way \cite{Belle05,BABAR10}.

The mass spectrum $D^{*0}\bar D^0 + c.c.$ in the $X(3872)\to
D^{*0}\bar D^0 + c.c.$ decay \cite{Belle10} looks as the typical
resonance threshold enhancement, see Fig. \ref{fig2}.

\end{multicols}
\ruleup
\begin{center}
\begin{tabular}{ccc}
\includegraphics[width=8.5cm,height=6cm]{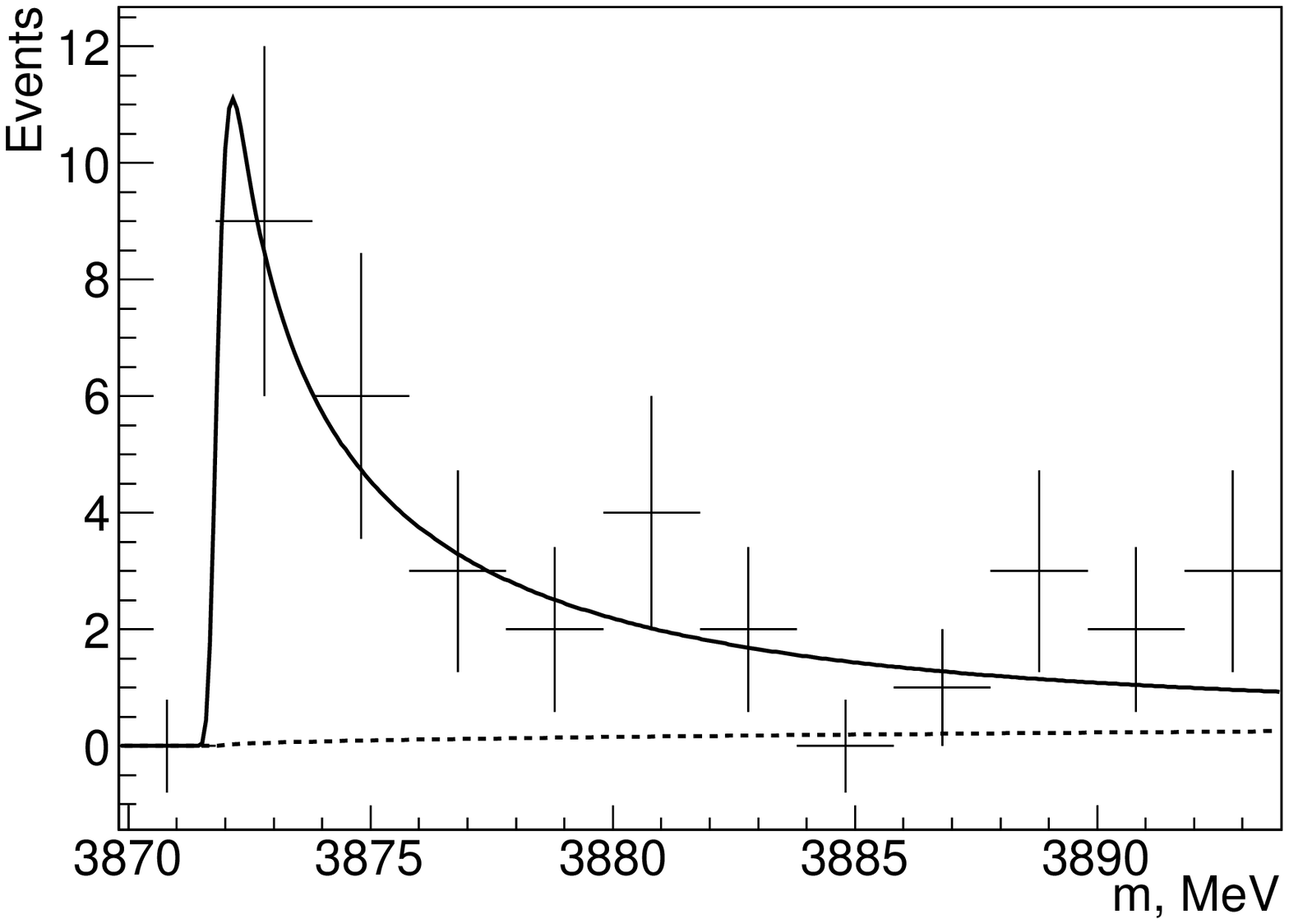}& \includegraphics[width=8.5cm,height=6cm]{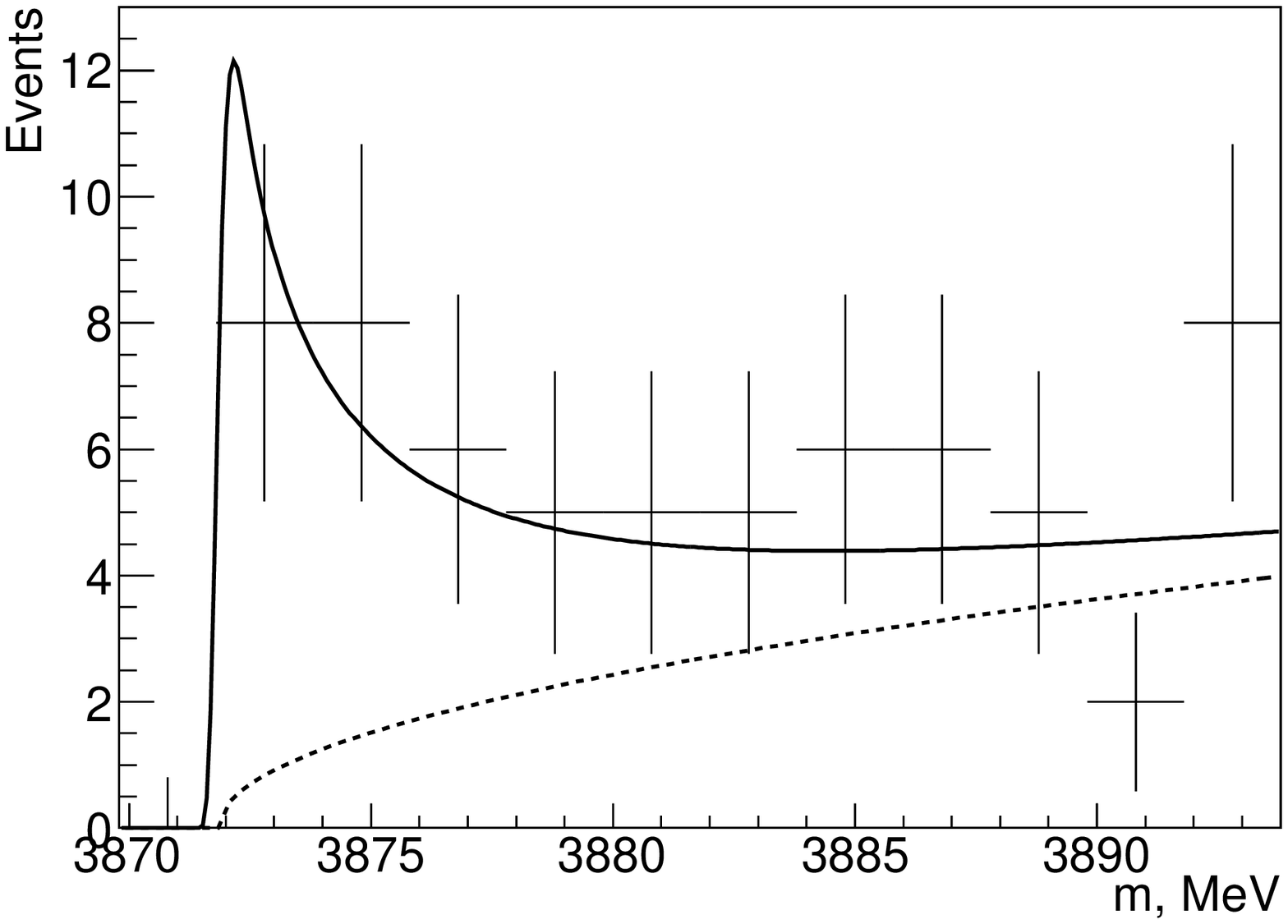}\\ (a)&(b)
\end{tabular}
\figcaption{The Belle data \cite{Belle10} on the invariant
$D^{*0}\bar D^0 + c.c.$ mass ($m$) distribution. The solid line is
our theoretical one with taking into account the Belle energy
resolution. The dotted line is a square root function for the
incoherent background. (a)  $D^{*0}\to D^0\pi^0$. (b) $D^{*0}\to
D^0\gamma$.  \label{fig2}}
\end{center}
\ruledown

\begin{multicols}{2}

 If structures in the above channels are manifestation of the same resonance, it is possible to  define the branching ratio
$BR(X(3872) \to D^{*0}\bar D^0 + c.c.)$ treating data  of the two
above  decay channels only.

We believe that the $X(3872)$ is the axial vector, $1^{++}$
\cite{LHCb,PDG14}. In this case the S wave dominates in the
$X(3872) \to D^{*0}\bar D^0 + c.c.$ decay and hence is described
by the effective Lagrangian
\begin{equation}
\label{L} L_{XD^{*0}D^0}(x)=g_AX^\mu\Bigl(D^0_\mu(x)\bar D^0(x)+
\bar D^0_\mu(x)D^0(x)\Bigr ).
\end{equation}

The width of the $X\to D^{*0}\bar D^0 + c.c.$ decay
\begin{equation}
\label{width}
 \Gamma(X \to D^{*0}\bar D^0 + c.c.\,,\, m
 )=\dfrac{g_A^2}{8\pi}\frac{\rho(m)}{m}\Biggl(1+\dfrac{{\bf k}^2}{3m_{D^{*0}}^2}\Biggr)\,,
 \end{equation}
where $\bf k$ is momenta of $D^{*0}$ (or $\bar D^0$) in the
$D^{*0}\,\bar D^0$ center mass system, $m$ is the invariant mass
of the $D^{*0}\,\bar D^0$ pair, $$ \rho(m)=\frac{2|{\bf k}|}{m}=
\frac{\sqrt{(m^2-m_+^2)(m^2-m_-^2)}}{m^2},\ m_\pm =m_{D^{*0}}\pm
m_{D^0}.$$

The second term in the right side of Eq. (\ref{width}) is very
small in our energy region and can be neglected. This gives us the
opportunity to construct the mass spectra for the  $X(3872)$
decays with the good analytical and unitary properties as in the
scalar meson case \cite{ads,nna-avk}.

The mass spectrum in the $D^{*0}\bar D^0 + c.c.$ channel
\begin{equation}
\label{SpectrumDD*} \dfrac{ dBR(X\to D^{*0}\bar D^0+c.c.\,,\, m
)}{dm} =4\dfrac{1}{\pi}\frac{m^2\Gamma(X\to D^{*0}\bar D^0,\,
m)}{|D_X(m)|^2}.
\end{equation}

The branching ratio of $X(3872) \to D^{*0}\bar D^0 + c.c.$
\begin{equation}
\label{BRDD*}
 BR(X\to
D^{*0}\bar D^0+c.c.) = 4\dfrac{1}{\pi}\int_{m_+}^\infty
\dfrac{m^2\Gamma(X\to D^{*0}\bar D^0,\, m)}{|D_X(m)|^2} dm\,.
\end{equation}

In others $\{i\}$ (non-$D^{*0}\bar D^0$) channels the $X(3872)$
state is seen as a narrow resonance that is why we write the mass
spectrum in the $i$ channel in the form
\begin{equation}
\label{Spectrumi}
 \dfrac{dBR(X\to i\,,\, m )}{dm}
=2\dfrac{1}{\pi}\frac{m_X^2\,\Gamma_i}{|D_X(m)|^2}\,,
\end{equation}
where $\Gamma_i$ is the width of the $X(3872)\to i$ decay.

The branching ratio of $X(3872)\to i$
\begin{equation}
\label{BRi}
 BR(X\to i) = 2\dfrac{1}{\pi}\int_{m_0}^\infty
\frac{m_X^2\Gamma_i}{|D_X(m)|^2} dm\,,
\end{equation}
where $m_0$ is the threshold of the $i$ channel.

  The inverse
propagator $D_X(m)$
 \begin{equation}
 \label{DX}
 D_X(m)= m_X^2-m^2 + Re(\Pi_X(m_X^2))-
\Pi_X(m^2)-\imath
 m_X\Gamma\,,
 \end{equation}
where $\Gamma=\Sigma\Gamma_i< 1.2$ MeV is the total width of the
$X(3872)$ decay into all non-($D^{*0}\bar D^0+c.c.$) channels.
\begin{equation}
 \label{PiX}
\Pi_X(s)=\dfrac{g_A^2}{8\pi^2}\left (I^{D^{*0}\bar
D^0}(s)+I^{D^{*+} D^{-}}(s)\right ),\ \ \ m^2=s\,.
\end{equation}

 When $m_+ = m_{D^*}+m_D\leq m$,   \vspace*{-3pt}
\begin{eqnarray}
 \label{ID}
&&  I^{D^*\bar D}(m^2)=
\dfrac{(m^2-m_+^2)}{m^2}\frac{m_-}{m_+}\ln\frac{m_{D^*}}{m_D}
\nonumber\\[3pt]
 &&  +\rho(m)\left [\imath\pi +
\ln\dfrac{\sqrt{m^2-m_-^2}-\sqrt{m^2-m_+^2}}{\sqrt{m^2-m_-^2}+\sqrt{m^2-m_+^2}}
\right ]\,
\end{eqnarray}
 where $m_-= m_{D^*}-m_D$\,.

  When  $m_- \leq m\leq m_+$,
\begin{eqnarray}
\label{m-mm+}
 && I^{D^{*}\bar D}(m)=
\dfrac{(m^2-m_+^2)}{m^2}\frac{m_-}{m_+}\ln\frac{m_{D^*}}{m_{D}}\nonumber\\[3pt]
 &&
-2|\rho(m)|\arctan\dfrac{\sqrt{m^2-m_-^2}}{\sqrt{m_+^2-m^2}}\,,
\end{eqnarray}
where $|\rho(m)|= \sqrt{(m_+^2-m^2)(m^2-m_-^2)}/m^2$.

When $m\leq m_-$ and $m^2\leq 0$,
\begin{eqnarray}
\label{mm-}
 && I^{D^*\bar D}(m)=\dfrac{(m^2-m_+^2)}{m^2}\frac{m_-}{m_+}\ln\dfrac{m_{D^*}}{m_{D}}\nonumber\\[3pt]
&&
-\rho(m)\ln\dfrac{\sqrt{m_+^2-m^2}-\sqrt{m_-^2-m^2}}{\sqrt{m_+^2-m^2}+\sqrt{m_-^2-m^2}}
\,.
\end{eqnarray}

Our branching ratios satisfy the unitarity
\end{multicols}
\begin{equation}
\label{UNITARITY}
 1= BR(X\to D^{*0}\bar D^0+c.c.)+BR(X\to
D^{*+}\bar D^-+c.c.)+ \sum_iBR(X\to i)\,.
 \end{equation}
 \begin{multicols}{2}

 Fitting the Belle data \cite{Belle11}, we take into account the
 Belle
 results \cite{Belle11}:
 $m_X= 3871.84\,\mbox{MeV}= m_{D^{*0}}+ m_{D^0}= m_+$ and
$\Gamma_{X(3872)}<1.2$ MeV  90\%CL,  that corresponds to $\Gamma
<1.2$ MeV,  which controls the width of the $X(3872)$ signal in
the $\pi^+\pi^- J/\psi(1S)$ channel and in every non-($D^{*0}\bar
D^0+c.c.)$
 channel. The results of our fit are  in the
 Table 1.
 \end{multicols}
Table 1.  $BR_{seen}$ is a branching ratio for $m\leq 3891.84$
MeV, $\Gamma$ in MeV, $g_A$ in GeV.\\[3pt]

\renewcommand{\tabcolsep}{4 mm}
\begin{tabular}{|cc|cccc|} \hline $\Gamma$ &
\hspace*{-6pt}$1.2_{-0.4}$ & mode & $D^{*0}\bar D^0+c.c.$ &
$D^{*+}D^{-}+c.c.$ & $Others$\\ \hline $g_A^2/8\pi$
 &\hspace*{-6pt} $1.4_{-1}^{+5}$ & $BR$ &
$0.6_{-0.1}^{+0.02}$ & $0.31_{-0.16}^{+0.13}$ &
$0.1_{-0.1}^{+0.3}$\\ \hline $\chi^2/Ndf$ &\hspace*{-6pt} 45/42 &
$BR_{seen}$ & $0.3_{-0.2}^{+0.1}$ & $0.03_{-0.02}^{+0.004}$ &
$0.09_{-0.1}^{+0.3}$\\ \hline
\end{tabular}
\begin{multicols}{2}

Our approach can serve as the guide in selection of theoretical
models for the $X(3872)$ resonance. Indeed, if $3871.68$ MeV
$<M_X<3871.95$ MeV and  $\Gamma_{X(3872)}=\Gamma <1.2$ MeV  then
for $g_A^2/4\pi<0.2$ GeV$^2$ $BR(X\to D^{*0}\bar D^0+c.c.\,; m\leq
3891.84\,\mbox{MeV})<0.3$. That is,  unknown decays of $X(3872)$
into non-$D^{*0}\bar D^0$ states are considerable or dominant.

For example, in Ref. \cite{Maiani} the authors considered
$m_X=3871.68$ MeV, $\Gamma=1.2$ MeV and
$g_{XDD^*}=g_A\sqrt{2}=2.5$ GeV, that is, $g_A^2/8\pi=0.1$
GeV$^2$. In this case $BR(X\to D^{*0}\bar D^0+c.c.)\approx 0.2$,
that is, unknown decays $X(3872)$ into non-($D^{*0}\bar D^0+c.c.)$
states are dominant. For details see Table 2.\\
\end{multicols}
Table 2. Branching ratios for the model from Ref. \cite{Maiani}.
$\Gamma$ in MeV, $m_X$ in MeV,  $g_A$ in GeV.\\[3pt]

\renewcommand{\tabcolsep}{4 mm}
\begin{tabular}{|cc|cccc|}
\hline $m_X$ & $3871.68$ & mode & $X\to D^{*0}\bar D^0+c.c.$ &
$X\to D^{*+}D^{-}+c.c.$ & $X\to Others$\\ \hline $\Gamma$ & $1.2$
& $BR$ & $0.176$ & $0.045$ & $0.779$\\ \hline $g_A^2/{8\pi}$ &
$0.1$ & $BR_{seen}$ & $0.14$ & $0.011$ & $0.761$\\ \hline
\end{tabular}
\vspace*{15pt}
\begin{multicols}{2}
\section{\bf\boldmath{$ $} $X(3872)$,
$I^G(J^{PC})=0^+(1^{++})$, as the $\chi_{c1}(2P)$ charmonium
\cite{NNAEVR15}}

Contrary to almost standard opinion that the $X(3872$) resonance
is the $D^{*0}\bar D^0+c.c.$ molecule or the $qc\bar q\bar c$
four-quark state, we discuss  the scenario where the $X(3872)$
resonance is the $c\bar c = \chi_{c1}(2P)$ charmonium which "sits
on" the  $D^{*0}\bar D^0$ threshold.

The two dramatic discoveries have generated a stream of the
$D^{*0}\bar D^0+D^0\bar D^{*0}$  molecular interpretations of the
$X(3872)$ resonance.

The mass of the $X(3872)$ resonance is 50 MeV lower than
predictions of the most lucky naive  potential models for the mass
of the $\chi_{c1}(2P)$ resonance,
\begin{equation}
\label{shiftmass} m_X-m_{\chi_{c1}(2P)}= -\Delta\approx -
50\,\mbox{MeV},
\end{equation}
and the relation between the branching ratios
\begin{eqnarray}
\label{isotopicviolation}
 && BR(X\to\pi^+\pi^-\pi^0J/\psi(1S))\nonumber\\[6pt]
 &&\sim BR(X\to\pi^+\pi^-J/\psi(1S))\,,
\end{eqnarray}
 that is interpreted as a strong
violation of isotopic symmetry.

  But the bounding energy is small,
$\epsilon_B<(1 \div 3)$ MeV. That is, the radius of the molecule
is large, $r_{X(3872)}> (3 \div 5)$ fm $ = (3 \div
5)\cdot10^{-13}$ cm. As for the charmonium, its radius is less one
fermi, $r_{\chi_{c1}(2P)}\approx 0.5$ fm $=0.5\cdot 10^{-13}$ cm.
 That is, the molecule volume is $100 \div 1000$ times as large
as the charmonium volume,
 $V_{X(3872)}/V_{\chi_{c1}(2P)} > 100 \div 1000$.

  How to explain sufficiently abundant inclusive production of the
rather extended molecule X(3872) in a hard process $pp\to X(3872)
+ anything$ with rapidity in the range 2,5 - 4,5 and transverse
momentum in the range 5-20 GeV \cite{LHCb12} ? Really,
\begin{eqnarray}
\label{pptoX(3872)} && \sigma (pp \to X(3872) + anything)
BR(X(3872)\to\pi^+\pi^-J/\psi)\nonumber\\[6pt]
&&=5.4\, \mbox{nb}
\end{eqnarray}
 and
\begin{eqnarray}
\label{pptopsi(2S)}
 && \sigma (pp\to\psi(2S) + anything)
BR(\psi(2S)\to\pi^+\pi^-J/\psi)\nonumber\\[6pt]
 &&=38\, \mbox{nb}.
\end{eqnarray}

But, according to Ref. \cite{PDG14}
\begin{equation}
\label{psi(2S)Jpsi}
 BR(\psi(2S)\to\pi^+\pi^-J/\psi)=0.34
\end{equation}
   while
\begin{equation}
\label{X(3872)Jpsi}
 0.023<BR(X(3872)\to\pi^+\pi^-J/\psi)<0.066
\end{equation}
according to Ref. \cite{Belle09}.

 So,
\begin{equation}
\label{pptoX(3872vspptopsi(2S)}
 0.74<\dfrac{\sigma (pp \to X(3872)
+ anything)}{\sigma (pp\to\psi(2S) + anything)}<2.1.
\end{equation}

 The extended molecule is produced in the hard process as intensively
as the compact charmonium. It's a miracle.

As for the problem of the mass shift, Eq. (\ref{shiftmass}), the
contribution of the $D^- D^{*+}$ and $\bar D^0 D^{*0}$ loops, see
Fig. \ref{fig3}, into the self energy of the $X(3872)$ resonance,
$\Pi_X(s)$, solves it easily. \vspace*{-60pt}
\begin{center}
\includegraphics[width=8cm,height=6cm]{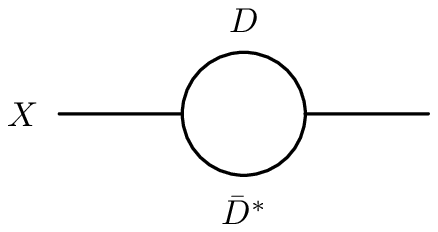}
\vspace*{-45pt} \figcaption{The contribution of the $\bar D^0
D^{*0}$ and  $D^- D^{*+}$ loops into the self energy of the
$X(3872)$ resonance. \label{fig3}}
\end{center}

 Let us  calculate
$I^{D^*\bar D}(s)$ in Eq. (\ref{PiX}) with help of a cut-off
$\Lambda$.
\begin{eqnarray}
 \label{ID}
&& I^{D^*\bar D}(s)=\int\limits_{m_+^2}^{\Lambda^2}
\dfrac{\sqrt{(s'-m_+^2)(s'-m_-^2)}}{s'(s'-s)}ds'\nonumber\\[3pt]
&&\approx 2 \ln\dfrac{2\Lambda}{m_+}
-2\sqrt{\dfrac{m_+^2-s}{s}}\arctan\sqrt{\dfrac{s}{m_+^2-s}}\,,
\end{eqnarray}
 where $ s<m_+^2\,,\ \ \Lambda^2\gg m_+^2$.

 The inverse propagator of the X(3872) resonance
\begin{equation}
\label{propagator}
 D_X(s)= m_{\chi_{c1}(2P)}^2-s -\Pi_X(s)-\imath
m_X\Gamma\,.
\end{equation}

The renormalization of mass
\begin{equation}
\label{renormalization}
 m_{\chi_{c1}(2P)}^2-m_X^2 -\Pi_X(m_X^2)=0
\end{equation}
results in
\begin{equation}
\label{Delta}
 \Delta \left (2m_X+\Delta\right
)=\Pi_X(m_X^2)\approx\left (g_A^2/8\pi^2\right
)4\ln(2\Lambda/m_+)\,.
\end{equation}

 If $\Delta = m_{\chi_{c1}(2P)} - m_X \approx
50$ MeV\,, then $g_A^2/8\pi\approx 0.2$ GeV$^2$ for $\Lambda=10$
GeV and $BR(X\to D^0\bar D^{*0}+ \bar D^0D^{*0})\approx 0.3$.
\footnote{The assumption of the determining role of the $D^*\bar D
+ c.c.$ channels in the shift of the mass of the $\chi_{c1}(2P)$
meson is based on the following reasoning. Let us imagine that $ D
$ and $ D ^ * $ mesons are light, for example, as the $ K $ and $
K ^ * $ mesons. Then the width of $ X (3872) $ meson is equal 50
MeV for $g_A^2/8\pi=0.2\, GeV^2$ that much more than the width of
its decay into all non-($D^{*0}\bar D^0$+c.c.) channels, $\Gamma <
1.2$ MeV. That is, in our case the coupling of the $X(3872)$ meson
with the $D^*\bar D + c.c.$ channels  is rather strong.}

Thus, we expect that a number of unknown mainly two-gluon  decays
of $X(3872)$ into non-$D^{*0}\bar D^0 + c.c.$ states are
considerable \footnote{Note that in the $\chi_{c1}(1P)$ case the
width of the two-gluon decays equals 0.56 MeV \cite{PDG14} that
agrees with $\Gamma<1.2$ MeV satisfactory.}.  The discovery of
these decays would be the strong (if not decisive) confirmation of
our scenario.

 As for $BR(X\to\omega J/\psi)\sim BR(X\to\rho
J/\psi)$, Eq. (\ref{isotopicviolation}), this could be a result of
dynamics. In our scenario the $\omega J/\psi$ state is produced
via the three gluons, see Fig. \ref{fig4}.
\begin{center}
\includegraphics[width=6.5cm,height=4cm]{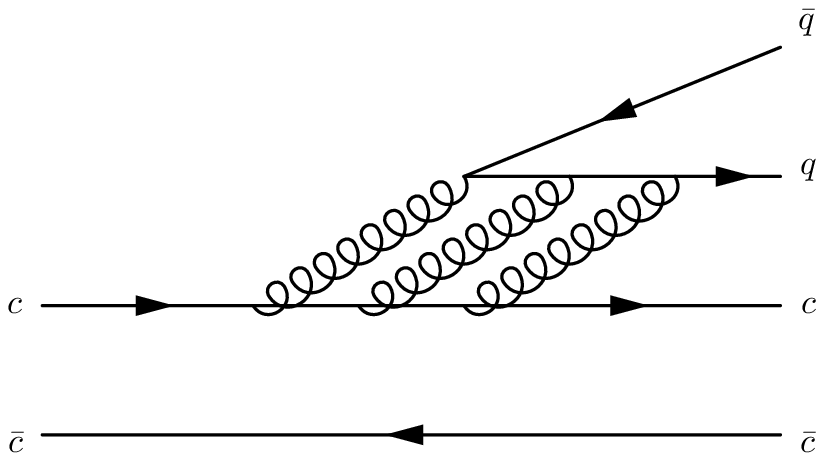}
\vspace*{-15pt}
\figcaption{The three-gluon production of the
$\omega$ and $\rho$ mesons (the $\rho$ meson via the contribution
$\sim m_u-m_d$ ).  All possible permutations of gluons  are
assumed.  \label{fig4}}
\end{center}
 As for the $\rho J/\psi$ state, it
is produced both via the one photon, see Fig. \ref{fig5}, and via
the three gluons (via the contribution $\sim m_u-m_d$ ), see Fig.
\ref{fig4}.
\begin{center}
\includegraphics[width=6.5cm,height=4cm]{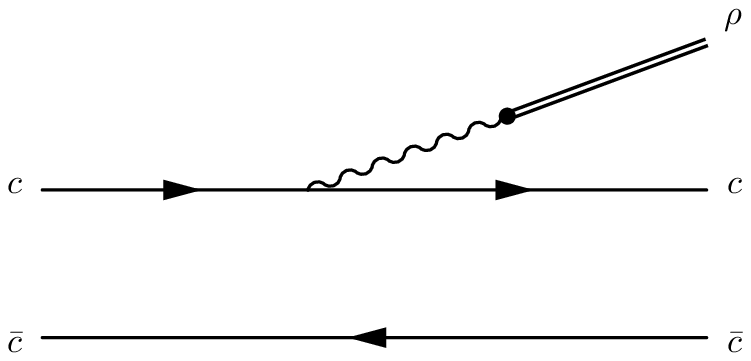}
\vspace*{-15pt}
 \figcaption{The one-photon production of the
$\rho$ meson. All possible permutations of  photon are assumed.
\label{fig5}}
\end{center}

 Close to our scenario is an example of the
$J/\psi\to\rho\eta'$ and $J/\psi\to\omega\eta'$  decays. According
to Ref. \cite{PDG14}
\begin{eqnarray}
\label{Jpsi}
 && BR(J/\psi\to\rho\eta')=(1.05\pm 0.18)\cdot10^{-4}\
\ \mbox{and}\nonumber\\
&& \ BR(J/\psi\to\omega\eta')=(1.82\pm
0.21)\cdot10^{-4}.
\end{eqnarray}

Note that in the $X(3872)$ case the $\omega$ meson is produced on
its tail ($m_X-m_{J/\psi}=775$ MeV), while the $\rho$ meson is
produced on a half.

It is well known that the physics of charmonium ($c\bar c$) and
bottomonium ($b\bar b$) is similar. Let us compare the already
known features of X(3872) with the ones of $\Upsilon_{b1}(2P)$.

Recently, the LHCb Collaboration published a landmark result
\cite{LHCb14}
\begin{equation}
\label{Xtogamma}
 \dfrac{BR(X\to\gamma \psi(2S))}{BR(X\to\gamma
J/\psi)}=C_X\left(\dfrac{\omega_{\psi(2S)}}{\omega_{J/\psi}}\right)^3=2.46\pm
0.7\,,
\end{equation}
where $\omega_{\psi(2S)}$ and $\omega_{J/\psi}$ are the energies
of the photons in the $X\to\gamma \psi(2S)$ and $BR(X\to\gamma
J/\psi)$ decays, respectively.

 On the other hand, it is known \cite{PDG14} that
 \begin{eqnarray}
 \label{chib1(2P)}
&& \dfrac{BR(\chi_{b1}(2P)\to\gamma
\Upsilon(2S))}{BR(\chi_{b1}(2P)\to\gamma
\Upsilon(1S))}=C_{\chi_{b1}(2P)}\left(\dfrac{\omega_{\Upsilon(2S)}}{\omega_{\Upsilon(1S)}}\right)^3\nonumber\\[6pt]
&&=2.16\pm0.28\,,
\end{eqnarray}
where $\omega_{\Upsilon(2S)}$ and $\omega_{\Upsilon(1S)}$ are the
energies of the photons in the $\chi_{b1}(2P)\to\gamma
\Upsilon(2S)$ and $\chi_{b1}(2P)\to\gamma \Upsilon(1S)$ decays,
respectively.

Consequently,
\begin{equation}
\label{CX}
 C_X=136.78\pm38.89
\end{equation}
and
\begin{equation}
\label{Cchib1(2P)}
 C_{\chi_{b1}(2P)}=80\pm10.37\
\end{equation}
as all most lucky versions of the  potential model predict for the
quarkonia $C_{\chi_{c1}(2P)}\gg 1$ and $C_{\chi_{b1}(2P)}\gg 1$.

 According to Ref. \cite{PDG14}
\begin{equation}
\label{omegaUpsilon(1S)}
 BR(\chi_{b1}(2P)\to\omega\Upsilon(1S))=\left
(1.63\pm^{0.4}_{0.34}\right )\%\,.
\end{equation}

If  the one-photon mechanism dominates in the $X(3872)\to\rho
J/\psi$ decay, see Fig. \ref{fig5}, then one should expect
\begin{eqnarray}
\label{onephotonrhoUpsilon(1S)}
 &&
BR(\chi_{b1}(2P)\to\rho\Upsilon(1S))\sim(e_b/e_c)^2\cdot
1.6\,\%\nonumber\\[3pt]
 &&=(1/4)\cdot 1.6\,\%= 0.4\%\,,
\end{eqnarray}
 where $e_c$ and
$e_b$ are the charges of the $c$ and $b$ quarks, respectively.

If  the three-gluon mechanism (its part $\sim m_u-m_d$ ) dominates
in the $X(3872)\to\rho J/\psi$ decay, see Fig. \ref{fig4}, then
one should expect
\begin{equation}
\label{threegluonrhoUpsilon(1S)}
 BR(\chi_{b1}(2P)\to\rho\Upsilon(1S))\sim 1.6\%\,.
\end{equation}

\section{Conclusion}

We believe that  discovery of a significant number of unknown
decays of $X(3872)$ into non-$D^{*0}\bar D^0 +c.c.$ states via two
gluons and discovery of the $\chi_{b1}(2P)\to\rho\Upsilon(1S)$
decay could decide destiny of X(3872).

Once more, we discuss the scenario where the $\chi_{c1}(2P)$
charmonium sits on the $D^{*0}\bar D^0$ threshold but not a mixing
of the giant $D^{*}\bar D$ molecule and the compact
$\chi_{c1}(2P)$ charmonium, see, for example, Refs. \cite{KR, BAK}
 and references cited therein. Note that the mixing of such
states requests the special justification. That is, it is
necessary to show that the transition of the giant  molecule into
the compact charmonium is considerable at insignificant
overlapping of their wave functions. Such a transition $\sim
\sqrt{V_{\chi_{c1}(2P)}/V_{X(3872)}}$ and a branching ratio of a
decay via such a transition $\sim V_{\chi_{c1}(2P)}/V_{X(3872)}$.

\section{Acknowledgments}

N. N. Achasov is grateful to Prof. Guangshun Huang
 for the offer to give the remote plenary talk.

\end{multicols}

\clearpage

\end{CJK*}

\vspace{10cm}

\begin{thebibliography}{90}

\vspace{2mm}

\bibitem{NNAEVR}
N. N. Achasov and E. V. Rogozina,  Pis'ma v ZhETF {\bf 100}, 252
(2014) [ JETP Letters {\bf 100}, 227 (2014)].
\bibitem{Belle11} S. K. Choi {\it et al.} (Belle Callaboration), Phys.
Rev. D {\bf 84}, 052004 (2011).
\bibitem{Belle05} K. Abe {\it et al.} (Belle Callaboration),
 arXiv:hep-ex/0505037.
\bibitem{BABAR10} P. del Amo Sanchez {\it et al.} (BABAR Callaboration), Phys.
Rev. D {\bf 82}, 011101 (R) (2010).
\bibitem{Belle10} T. Aushev {\it et al.} (Belle Callaboration), Phys.
Rev. D {\bf 81}, 031103 (2010).
\bibitem{LHCb} R. Aaij {\it et al.} (LHCb Callaboration), Phys. Rev. Lett. {\bf
110}, 22001 (2013).
\bibitem{PDG14}  K. A. Olive {\it et al.} (Particle Data Group),
Chin. Phys. C {\bf 38}, 090001 (2014).
\bibitem{ads} N. N. Achasov, S. A. Devyanin,  and  G. N. Shestakov, Yad. Fiz. {\bf
32}, 1098 (1980) [ Sov. J. Nucl. Phys. {\bf 32}, 566 (1980)].
\bibitem{nna-avk} N.N. Achasov and A.V. Kiselev, Phys. Rev. D {\bf
70}, 111901 (R) (2004).
\bibitem{Maiani} L. Maiani {\it et al.}, Phys. Rev. D {\bf 87},
111102 (R) (2013).
\bibitem{NNAEVR15}  N. N. Achasov and  E. V. Rogozina, Mod. Phys. Lett.
A {\bf 30}, 1550181 (2015).
\bibitem{LHCb12} R. Aaij {\it et al.} (LHCb Collaboration),  Eur. Phys.
J. C {\bf 72}, 1972 (2012).
\bibitem{Belle09}  C.-Z. Yuan (Belle Collaboration), Proceedings of the XXIX PHYSICS IN
COLLISION, Kobe, Japan, 2009, arXiv: hep-ex/0910.3138.
\bibitem{LHCb14} R. Aaij {\it et al.} ( LHCb Collaboration), Nucl.
Phys. B {\bf 886}, 665 (2014).
\bibitem{KR} M. Karliner and J. L. Rosner, Phys. Rev. D {\bf 91},
014014 (2015).
\bibitem{BAK} M. Butenschoen, Z.-G. He, and B. A. Kniehl, Phys. Rev. D {\bf 88}
011501 (R) (2013).


\end{thebibliography}
\end{document}